\documentstyle[jjc,epsfig]{article}
\font\tencop=bbm10
\font\sevencop=bbm7 \font\fivecop=bbm5
\newfam\copfam\scriptscriptfont
\copfam=\fivecop\textfont\copfam=\tencop\scriptfont\copfam=\sevencop

\begin{document}

\title{Aspects non-perturbatifs des th\'eories de jauge 
(supersym\'etriques)}
\author{Frank Ferrari\\
\it Laboratoire de Physique Th\'eorique de l'\'Ecole
Normale Sup\'erieure\\
\it 24 rue Lhomond, 75231 Paris Cedex 05\\
\tt Frank.Ferrari@lpt.ens.fr\\}

\abstract{\rightskip=1.5pc \leftskip=1.5pc
\`A un niveau \'el\'ementaire, on pr\'esente certains aspects 
non-perturbatifs des th\'eories de jauge non-ab\'eliennes \`a quatre
dimensions d'espace-temps. Des 
r\'esultats rigoureux ont pu \^etre obtenus dans le cadre des 
th\'eories supersym\'etriques, mettant en \'evidence une physique 
tr\`es riche li\'ee \`a la dynamique des champs de jauge en 
couplage fort. Les nouveaux ph\'enom\`enes qui sont apparus joueront 
peut-\^etre un r\^ole dans l'\' elaboration des mod\`eles
ph\'enom\'enologiques du futur. N\' eanmoins, l'application 
quantitative de ces id\' ees \` a des th\' eories non-supersym\' 
etrique, telle que la chromodynamique quantique, reste hors de port\' ee
pour le moment.}
\maketitle
\setbox1=\hbox{LPTENS-97/01}
\setbox2=\hbox{\tt hep-ph@xxx/9702399}
\setbox3=\vtop{\null\epsfbox{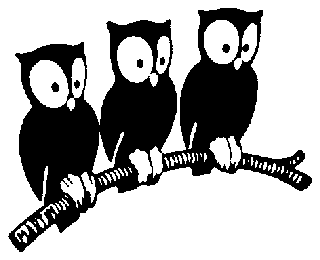}}
\makeatletter
\global\@specialpagefalse
\def\@oddhead{\hskip -.5cm\vbox{\vskip 1.4cm \box3}\hfill\vbox{\vskip -1.4cm
\box1\box2\vskip 1cm}}
\let\@evenhead\@oddhead
\vskip 5mm
Les th\' eories que nous consid\' erons partagent de nombreuses caract\' 
eristiques avec les th\' eories r\' ealistes \`a la base du mod\` ele 
standard. Principalement, elles contiennent des champs de jauge 
non-ab\'eliens,
\'eventuellement coupl\' es \` a la mati\` ere, et sont asymptotiquement 
libres. De plus, il existe des corrections non-perturbatives \` a la s\' 
erie de perturbation. Semi-classiquement, ces corrections sont dues 
aux instantons, qui sont pr\' esents dans toutes les th\' eories de 
jauge non-ab\' eliennes.
La sym\' etrie de jauge est spontan\'ement bris\'ee par la valeur 
moyenne non-nulle d'un champ scalaire complexe
$\phi $, qui appartient \` a la repr\' 
esentation adjointe du groupe de jauge, et qui joue le r\^ ole du 
boson de Higgs. G\' en\' eriquement, le groupe de jauge $G$ est bris\' 
e en $U(1)^{r}$ o\` u $r$ est le rang du groupe.
Je voudrais souligner de plus quatre caract\' eristiques particuli\` eres 
des th\' eories supersym\'etriques $N=2$, qui sont celles pour 
lesquelles des r\' esultats quantitatifs rigoureux peuvent \^ etre 
obtenus pour le spectre de masse et le contenu en particules,
et qu'il est bon de garder \`a l'esprit.
Tout d'abord, il n'existe pas de fermions chiraux. Ensuite, la s\' 
erie de perturbation s'arr\^ ete \` a une boucle; il n'y a donc aucun 
probl\`eme de resommation, et l'on peut consid\' erer sans 
difficult\' e d'\'eventuelles corrections non-perturbatives.
Troisi\` emement, ces corrections non-perturbatives peuvent \^ etre 
calcul\' ees semi-classiquement sans ambiguit\' e; les probl\` emes 
infrarouges venant de l'int\' egration sur la taille de l'instanton, 
bien connus en QCD, sont parfaitement ma\^ \i tris\' es ici.
Enfin, le potentiel scalaire est du type $\mathrm{tr}\, [\phi ,\phi 
^{\dagger}]^{2}$ et a donc des directions plates, qui ne sont pas lev\' 
ees au niveau quantique. Ceci veut dire que la valeur moyenne du Higgs 
n'est pas d\'etermin\'ee mais doit \^ etre consid\' er\' ee 
comme un param\`etre suppl\'ementaire de la th\'eorie. 
Sa donn\' ee fixe une \' echelle d'\' energie, \` 
a comparer \` a l'\'echelle $\Lambda $ g\' en\' er\' ee 
dynamiquement en raison de la libert\' e asymptotique.
Le r\' egime de couplage fort est celui pour lequel 
$\vert\langle\phi\rangle\vert $ est de l'ordre, ou plus petit, que
$\Lambda $.

Une question de physique \'evidente \`a
poser une fois le lagrangien d'une th\' eorie connu est la suivante: 
quelles sont les particules et leurs caract\' eristiques (masse, spin,
etc\dots ) qu'un exp\' erimentateur pourra \'eventuellement d\'etecter 
dans un monde d\' ecrit par la th\' eorie \' etudi\' ee? 
On ne peut r\' epondre en g\' en\' eral \`a cette question en lisant 
simplement le lagrangien. M\^ eme \` a un niveau perturbatif,
de gros efforts ont \' et\' e n\' ecessaires
dans le pass\' e pour comprendre comment une th\' eorie de jauge 
non-ab\'elienne pouvait contenir des particules {\em massives} de spin 1
et \^etre renormalisable.
\`A un niveau non-pertur\-ba\-tif, des ph\'enom\`enes spectaculaires sont 
susceptibles de se produire. L'un d'eux est le confinement des quarks, 
attendu en QCD.
De plus, il existe dans nos th\' eories, comme dans le mod\` ele de 
Georgi-Glashow, des mono\-p\^o\-les magn\' etiques et des dyons 
(particules \` a la fois charg\' ees \'electriquement et magn\' 
etiquement), en plus des par\-ti\-cu\-les habituelles du spectre perturbatif
(photon, bosons W, etc\dots). Malgr\'e l'origine tr\`es diff\'erente 
de toutes ces particules, il existe une formule remarquable, que l'on 
peut d\'eduire facilement dans le mod\`ele de Georgi-Glashow \`a un 
niveau semi-classique, qui donne la masse d'une particule quelconque 
du mod\`ele en fonction de la valeur moyenne du Higgs, de la charge 
\'electrique $Q_{e}$ et de la charge magn\'etique $Q_{m}$:
\begin{equation}
\label{BPSmass}
M=|\langle\phi\rangle |\, \sqrt{Q_{e}^{2}+Q_{m}^{2}}.
\end{equation}
Cette formule, invariante par \'echange de $Q_{e}$ et $Q_{m}$,
sugg\`ere qu'une \'equivalence entre particules charg\'ees 
\'elec\-tri\-quement et particules charg\'ees magn\'eti\-que\-ment 
pourrait exister [1]. Une telle \'equivalence, si elle \'etait correcte, 
aurait des cons\'equences remarquables. En effet, si $g$ est la 
constante de couplage de jauge de notre th\'eorie, la charge 
\'electrique $Q_{e}$ est bien s\^ur proportionnelle \`a $g$, mais la 
charge magn\'etique $Q_{m}$ doit \^etre proportionnelle \`a $1/g$, un 
r\'esultat d\^u \`a Dirac. Ainsi, si le couplage \'electrique $g$ est 
grand (comme en QCD), et que les d\'eveloppements perturbatifs 
habituels ne sont pas valables, il est envisageable de consid\'erer une 
formulation de la th\'eorie en termes de particules charg\'ees 
magn\'etiquement, qui elles ont un couplage en $1/g$ qui est faible et 
en terme duquel on peut donc \'ecrire des d\'eveloppements perturbatifs.

Comme on peut s'y attendre, une telle image ne peut \^etre correcte 
en toute g\'en\'eralit\'e. Tout d'abord, la formule fondamentale 
(\ref{BPSmass}) ne r\'esiste pas aux corrections quantiques, {\em \`a 
moins de consid\'erer des th\'eories ayant au moins deux
supersym\'etries.} De plus, une \'equivalence quantique compl\`ete entre 
deux th\'eories de jauge ayant des couplages inverses
ne peut \^etre vraie que si la fonction $\beta $ est nulle dans les deux 
th\'eories, ce qui limite fortement les possibilit\'es. Notons tout de 
m\^eme que de tels cas id\'eaux existent et ont un int\'er\^et 
th\'eorique ind\'eniable. N\'eanmoins, dans la plupart des cas, on 
peut au mieux esp\'erer que la dualit\'e \'electrique-magn\'etique 
soit vraie de mani\`ere approch\'ee, dans certaines limites, comme au 
niveau de l'action effective \`a basse \'energie. Par exemple, 
lorsque le groupe de jauge est bris\'e en ${\mathrm U(1)}^r$, l'action 
effective est une th\'eorie de jauge {\em ab\'elienne} dont la 
constante de couplage $g_{\mathrm eff}$ tend vers zero dans 
l'infrarouge et est donc susceptible d'\^etre reli\'ee \`a la 
constante de couplage $g$ de la th\'eorie microscopique 
asymptotiquement libre par une relation du type $g_{\mathrm eff}\sim 
1/g$.

Comment donc, au-del\`a de ces consid\'erations qualitatives, 
d\'ecider si une th\'eorie a une chance d'\^etre duale d'une autre
(au moins dans certaines limites)? 
On peut montrer qu'une telle
\'equivalence \`a des cons\'equences hautement non-triviales sur le 
spectre des particules d'une th\'eorie donn\'ee. Par exemple, les 
dyons susceptibles d'\^etre \'echang\'es avec les 
particules charg\'ees \'elec\-tri\-que\-ment doivent, au-del\`a de leur 
masse, avoir des nombres quantiques (spin, saveur, \dots) compatibles. 
Ceci est bien possible dans le cadre des th\'eories super\-sym\'e\-tri\-ques.
D'autre part, ces dyons sont en g\'en\'eral des 
{\em \'etats li\'es} de dyons \'el\'ementaires. L'existence 
m\^eme de ces \'etats li\'es au niveau quantique est loin d'\^etre \'evidente.
Pour l'\'etablir, il faut mettre au point une th\'eorie quantique des 
monop\^oles (qui est une m\'ecanique quantique supersym\'etrique), dont 
l'\'etude constitue un probl\`eme math\'e\-ma\-ti\-que tr\`es complexe. La 
difficult\'e de ces probl\`emes est telle que lorsque Sen montra en 
1994 l'existence d'\'etats li\'es \`a deux monop\^oles dans la 
th\'eorie la plus simple (ayant quatre supersym\'etries) [2],
alors que la 
dua\-li\-t\'e implique en fait l'existence d'\'etats li\'es avec un nombre 
quelconque de monop\^oles, il se d\'eclencha un v\'eritable engouement 
pour ces sujets, qui ne s'est pas \'eteint depuis, et dont l'un des
ach\`evements les plus remarquables reste 
le d\'etermination exacte par Seiberg
et Witten de l'action effective \`a basse \'energie dans certains
mod\`eles [3].

Il est important de noter que les d\'eveloppements r\'ecents, s'ils
sont confin\'es (pour l'instant?) aux th\'eories supersym\'etriques,
ont d'ores et d\'ej\`a jet\'e une lumi\`ere nouvelle sur la th\'eorie
des champs et en particulier sur les ph\'enom\`enes attach\'es \`a
la dynamique des champs de jauge en couplage fort.
Je terminerai cet expos\'e en illustrant ceci par trois exemples.
Le premier concerne le statut de la particule \'el\'ementaire.
Depuis toujours, la particule \'el\'ementaire a \'et\'e d\'efinie
comme \'etant un constituant {\em sans structure interne}.
Or, j'ai soulign\'e ci-dessus que la dualit\'e pr\'edisait
l'\'equivalence entre une formulation des th\'eories
en termes des particules \'el\'ementaires habituelles (charg\'ees
\'electriquement) et une formulation en termes
de monop\^oles magn\'etiques ou de dyons {\em qui apparaissent
dans la formulation initiale de la th\'eorie comme \'etant des
\'etats li\'es.} Ainsi, le statut de particule sans structure interne
ou composite est-il relatif, et d\'epend du point de vue (de la formulation
de la th\'eorie) adopt\'e. Ceci constitue \`a mes yeux une r\'evolution
conceptuelle de grande importance.
Le deuxi\`eme exemple concerne le statut des bosons de jauge massifs.
\`A la brisure spontan\'ee de la sym\'etrie de jauge est associ\'ee
habituellement la pr\'esence de particules massives de spin 1,
les ``bosons W:'' c'est le m\'ecanisme de Higgs. Cependant, la preuve
de l'existence de ces bosons W est de nature perturbative 
(contrairement, par exemple, au th\'eor\`eme de Goldstone),
et sa validit\'e pour des th\'eories fortement coupl\'ees peut \^etre
questionn\'ee. En fait, il a \'et\'e d\'emontr\'e pour la premi\`ere fois
dans [4] que dans certains r\'egimes de la QCD supersym\'etrique,
aucune particule massive de spin 1 n'\'etait associ\'ee \`a la brisure
de la sym\'etrie de jauge SU(2) en U(1). Le spectre de la th\'eorie
est en fait compl\`etement diff\'erent du contenu en champ dans le
lagrangien, ce qui montre qu'une formulation lagrangienne habituelle
n'est pas du tout adapt\'ee \`a la physique du couplage fort.
Pourra-t-on trouver une formulation mieux adapt\'ee? Ceci nous am\`ene
\`a mon troisi\`eme et dernier exemple, qui se situe \`a un niveau plus
sp\'eculatif, et qui concerne le lien entre la th\'eorie des cordes et 
les th\'eories de jauge. Il est apparu que certains des aspects
curieux de la physique des champs de jauge en couplage fort avaient une
interpr\'etation tr\`es naturelle en th\'eorie des cordes.
Cette th\'eorie, au-del\`a de son int\'er\^et intrins\`eque
en tant que th\'eorie candidate de la gravit\'e quantique,
pourrait fournir un cadre adapt\'e \`a l'\'etude des ph\'enom\`enes
non-perturbatifs en th\'eorie des champs, un peu comme les th\'eories
conformes ont permis de comprendre les ph\'enom\`enes critiques
bidimensionnels [5]. 

\noindent {\bf R\'ef\'erences}

\noindent {[1]} C. Montonen et D. Olive, Phys. Lett. B72 (1977)\\
\hphantom{[1]} 117.\\
{[2]} A. Sen, Phys. Lett. B329 (1994) 217.\\
{[3]} N. Seiberg et E. Witten, Nucl. Phys. B426 (1994)\\
\hphantom{[1]} 19;
erratum B430 (1994) 485,\\
\hphantom{[1]} N. Seiberg et E. Witten, Nucl. Phys. B431 (1994) \\
\hphantom{[1]} 484.\\
{[4]} F. Ferrari et A. Bilal, Nucl. Phys. B469 (1996) 387,\\
\hphantom{[1]} A. Bilal et F. Ferrari, Nucl. Phys. B480 (1996) 589.\\
{[5]} E. Witten, IASSNS-95-63, hep-th/9507121.
\end{document}